\documentclass[useAMS,usenatbib]{mn2e}
\usepackage{graphicx}
\usepackage{epstopdf}

\title[Time dependent gamma ray burst selection effects]{GRB optical afterglow and redshift selection effects: The learning curve effect at work}
\author[D. M. Coward]{D. M. Coward$^{1}$\thanks{E-mail:
coward@physics.uwa.edu.au}\\
$^{1}$School of Physics, University of Western Australia, M013, Crawley WA 6009, Australia\\
}
\begin{document}

\date{Accepted Received ; in original form }

\pagerange{\pageref{firstpage}--\pageref{lastpage}} \pubyear{2008}

\maketitle

\label{firstpage}

\begin{abstract}
We show how the observed gamma ray burst (GRB) optical afterglow (OA) and redshift distributions are changing in time from selection effects. For a subset of {\it Swift} triggered long duration bursts, we show that the mean time taken to acquire spectroscopic redshifts for a GRB OA has evolved to shorter times. We identify a strong correlation between the mean time taken to acquire a spectroscopic redshift and the measured redshift. This correlation reveals that shorter response times favour smaller redshift bursts. This is compelling evidence for a selection effect that biases longer response times with relatively brighter high redshift bursts. Conversely, for shorter response times, optically fainter bursts that are relatively closer are bright enough for spectroscopic redshifts to be acquired. This selection effect could explain why the average redshift, $<z>\approx2.8$ measured in 2005, has evolved to $<z>\approx2$, by mid 2008. Understanding these selection effects provides an important tool for separating the contributions of intrinsically faint bursts, those obscured by host galaxy dust and bursts not seen in the optical because their OAs are observed at late times. The study highlights the importance of rapid response telescopes capable of spectroscopy, and identifies a new redshift selection effect that has not been considered previously, namely the response time to measure the redshift. 
\end{abstract}

\begin{keywords} gamma-rays: bursts
\end{keywords}

\section{Introduction}

GRBs \footnote {Hereafter GRB refers to bursts classified as long.} are extremely bright transients observed out to great cosmological distances. Because of their high luminosity in $\gamma$-rays, they are a unique probe to their host galaxies in the high-$z$ Universe. At least a fraction of GRBs have been associated with the collapse of massive stars via the association of supernova signatures observed with the fading GRB optical afterglow e.g. \citep{hjorth03,stan03}. The afterglow most likely originates from an external shock produced as the blast wave from the GRB collides with the interstellar medium (ISM).  The successive afterglows at progressively longer wavelengths (X-ray, optical, radio) are a result of the expanding shock wave interacting with the ISM, causing it to slow down and lose energy. The GRB-supernova connection implies that GRBs should track the star formation rate (SFR) of massive stars, and could be used as a complementary probe of the SFR in the high-$z$ regime where optical data is lacking.  

With the launch of NASAs {\it Swift} satellite in 2004 November, a new era of rapid GRB localization was born. A slew of detectors on board {\it Swift} provided the means to rapidly find GRBs with small error boxes, enabling rapid follow up of the optical afterglow by dedicated ground-based telescopes. Prior to {\it Swift}, only about 50\% of localized GRBs were identified with an optical afterglow.  The high sensitivity of {\it Swift} coupled with the growing number of rapid response ground-based telescopes capable of spectroscopy promised to fill the gaps. Surprisingly, this did not happen: optical/NIR afterglows have been found for nearly $80\%$ of GRBs but only 40--50\% have measured redshifts \citep{T2007}.  It is now established that some GRBs may not have an afterglow and are defined as `dark' bursts. Jakobsson et al. (2004) define a dark burst by suggesting that the spectral slope between the optical and X-ray is $\beta_{ox} < 0.5$. Fynbo et al. (2008) point out that up to 25\% of Swift bursts fulfill this condition--see also \cite{Rol05}.

\section{Dark GRBs and selection effects}
We are now confronted with another problem -- the issue of the missing GRB optical afterglows. 
\cite{floch06} show out that {\it Spitzer} observations of long GRB are not associated
with the massive and luminous infrared galaxies that are associated with the bulk of star
formation in the early Universe. This work supports claims arguing for a GRB host population dominated by blue, young and low-mass galaxies and that dust extinction would not play a prominent role in such GRB hosts. However, there is some evidence that dust extinction in the local ISM of the host along the line of sight is the cause of some of the dark bursts. Evidence for this is shown in the study of the host galaxy of GRB 030115 \citep{lev06}. Its optical afterglow was fainter than many upper limits for other bursts, suggesting that without early NIR observations it would have been classified as a dark burst. Both the colour and optical magnitude of the afterglow probably arise from dust extinction and indicate that at least some optical afterglows are very faint due to dust along the line of sight. We must conclude that at least some fraction of dark bursts are caused by dust extinction in the host galaxy .

Roming et al. (2006), used very early observations of {\it Swift} GRBs to investigate the most probable cause of optically dark GRBs. They find that $\sim 25\%$ of the afterglows in their sample are 
extincted by Galactic dust, $\sim 25\%$ are obscured by absorption in the immediate GRB environment and $\sim 30\%$ are most likely attributable to Ly-alpha blanketing and absorption at high redshift. So it is highly likely that the dark bursts result from a combination of extinction factors relating to the GRB environment, host galaxy type and galaxy redshift  \citep{MR03}. Schady et. al. (2007) show that many OAs have significant optical extinctions compared to the optically bright bursts and argue that extinction could account for many of the OAs not observed by the Ultra-Violet Optical Telescope onboard {\it Swift}.
In support of this, \cite{T2008} show that dust is the likely cause of the colour and faintness of the red OA of GRB 060923A. Furthermore, \cite{JAUN08} show that the highly extinguished afterglow of GRB 070306 is due to host galaxy dust extinction.

The probability of obtaining a reliable GRB redshift is directly related to the signal-to-noise ratio of the absorption or emission lines. Ideally, multiple `strong' lines are required, but this is made difficult because GRB OA brightness decays rapidly in time. Combined with the fact that many GRB host galaxies are too faint for redshifts to be obtained, the time taken to image the OA with medium to large telescopes becomes a critical factor. This was first pointed out by \cite{fiore07} for the observed discrepancy between the {\it HETE} and {\it BeppoSAX} redshift distributions compared to {\it Swift}.     

It is clear that these optical afterglow selection effects will propagate through to the observed redshift distribution. In addition to afterglow extinction biases above, the so-called `redshift desert' in $z\approx1-2$ is a region where it is difficult to measure redshifts because of the lack of strong emission lines \citep{fiore07}. However, we note that for absorption lines, the Mg II doublet is prominent in $0.4 < z < 2.2$, so that the redshift desert may not play such a prominent role in redshift determination.

These problems were further highlighted by \cite{Cow07} and \cite{Cow08}. They argued that in $z = 0-1$, the GRB redshift distribution should increase rapidly because of increasing differential volume sizes and strong SFR evolution. Until mid 2007, this characteristic in the {\it Swift} redshift distribution was not apparent. To account for this discrepancy, they argue that other biases, independent of the {\it Swift} sensitivity, are required. The lack of measured redshifts in $z \approx 1-2$ up to 2007, discussed by \cite{Cow08}, is surprising and has not been resolved.

\subsection{GRB redshifts and the `Learning Curve'}
Assuming a redshift can be measured from any burst in the Universe and that GRBs are independent events (see Howell et al. 2007), then the bursts will follow a Poisson distribution defined by their mean rate observed in our frame. Furthermore, the mean GRB redshift will fluctuate about a constant assuming an observation duration much less than any cosmological epoch. This also implies that the observed time-series comprised of GRB redshifts will be stationary in time i.e. the statistical moments will not drift in time. Small number statistics will obviously  result in scatter of the statistical moments, but cannot be the cause of any evolution in the statistical moments of the GRB redshift distribution on a time-scale of years.

A shift of the mean of the GRB redshift distribution was observed in the early part of the {\it Swift} mission \citep{berg05}. This was attributed to the improved sensitivity and more accurate localisation by {\it Swift} resulting in a bias for fainter and higher redshift bursts. \cite{Jak06} showed that within the first year of {\it Swift} the mean redshift for a subset of 28 bursts had drifted to about 2.8, about double that of the pre-{\it Swift} average redshift. 

Assuming that satellite sensitivity is the dominant factor for determining redshift statistics, one could assume that the high mean redshift observed in the early part of the mission would remain fairly constant. If other factors impact on the statistics, such as the time taken to acquire high signal-to-noise spectra, the statistics may well reflect this. Given that {\it Swifts} sensitivity and GRB localisation ability has not degraded over time, one must consider the next link in the chain for measuring GRB redshifts: optical follow up by telescopes capable of spectroscopy. 

Given the above argument, any gross time-dependent evolution observed in the mean of the GRB redshift distribution must result from how the redshifts are measured. As discussed above, this is not a straightforward process, but depends on obtaining high quality spectra from OAs that are decaying rapidly in time. We will show in this {\it Letter} that the time taken to measure a spectroscopic redshift can cause an important selection effect. By analysing the delay time for spectroscopic redshift data as time-series, we show how the non-stationarity of the data can be used to constrain the origin of biases that plague this data. This type of analysis can help discriminate between ground-based optical selection effects and possible intrinsic characteristics of GRBs.

The learning curve effect,  first mathematically modelled by \cite{W1936}, has a strong influence in any large organization that is technology and skill dependent. It is based on the idea that efficiency for a particular task improves with experience and technology. It has become an important tool in management to model how productivity and efficiency improves in large organisations \citep{Adler91}. 

With the globalization of optical astronomical techniques and knowledge, the learning curve effect is expected to influence the efficiency of targeted transient surveys, such as supernova search programs (Coward 2008 in preparation). For the case of GRB OA spectroscopic redshift measurements, the effect is very subtle but nonetheless important. This basic idea can be related to GRB optical follow-up in terms of the efficiency of localizing, imaging and obtaining high quality spectra suitable for measuring redshift. In particular, a critical factor that determines the efficiency of these tasks is the time it takes to optically localize the rapidly fading OA and to obtain high signal-to-noise spectroscopy. One would expect that this task should be improving in time and we seek to identify if this effect is impacting on the redshift distribution statistics as a function of time.  

\section{Data analysis and results}
We select 192 GRBs detected by {\it Swift} from 2005 March to 2008 July with high-energy emission duration greater than 4 s. After removing those bursts without redshifts, the remaining 91 consist of those with absorption and emission spectra of the OA. From this population, we select a subset of 64 from GCN circulars that have response times from when the burst was triggered by {\it Swift's} BAT to the acquisition of a spectroscopic redshift. We have attempted to excluded burst redshifts measured from the host galaxy at very late times as obtaining these spectra did not depend critically on the response time of the telescope. 

We employ two additional statistics for the analysis: the optical brightness at discovery and the time of the burst. The optical brightness is not measured with the same filters across the data. Nonetheless it is a useful parameter for investigating if the efficiency of localizing bursts by rapid follow up of optical telescopes is changing in time. The redshift data above are analysed as time series to probe how the statistical moments--i.e. the mean, variance and the discovery rate--evolve over the mission time of {\it Swift}. To determine if the data is non-stationary in time, a moving average filter spanning 4 nearest neighbours on each side of an event is employed.

Fig. 1 plots the raw redshifts and output from the moving average filter (using 4 nearest neighbours) as a time-series. The associated GRB OA magnitudes at discovery are also plotted.  Although there are fluctuations over periods of months, there is a clear downward trend in the redshift averages over a 3 year period. We test if the observed non-stationarity of the redshift time-series is related to how the redshifts are measured, in particular the time taken to obtain spectroscopic redshifts.   

Fig. 2 plots both the response times to acquire these 64 redshifts and a moving average of this data, $<T_z>$, against the time when the burst occurred.
The plots shows a definite long term trend in the response times. To test if $<T_z>$ is affecting the selection of GRBs in a certain distance range, Fig. 3 plots $<T_z>$ against the mean redshift. The plot provides compelling evidence that there is a significant selection effect at work. There appears to be a strong correlation between the response times to acquire a redshift and the redshift measured via spectroscopy of the OA. 

Interpreting this result provides the key to understanding the observed drift in the mean GRB redshift over time. It is  clear that the probability of observing smaller GRB redshifts has increased up to mid-2008. From Fig. 3, the shortest waiting times to acquire a redshift  correspond, on average, to smaller redshift bursts. This observation can be understood if these bursts have rapidly fading OAs that would have been too faint for high quality spectroscopy if observed at later times. Conversely, the longer duration response times preferentially select those bursts with very bright OAs that can be observed spectroscopically at higher redshift. Furthermore, this selection effect can be reconciled with the previously observed lack of redshifts in $z<2$, as noted by several studies including Coward et al. (2008). 

Fig. 4 plots the cumulative growth in measured redshifts as a function of observation time for the redshift regimes $z>2$ and $z<2$. The plots show that the number of high-$z$ bursts measured is roughly constant, compared to the marginally non-linear increasing rate of the $z<2$ events. It is apparent that the probability of measuring a $z<2$ burst is increasing with time, a result supporting the observed drift to a smaller mean redshift shown in Figures 1 and 3. The root of this selection effect may originate from intrinsic differences in the OA brightness (at the source) between the low-medium and high redshift bursts. 

\section{Summary and discussion}
The results based on the analysis are summarized below:
\newline 1. A time-dependent trend in the mean of the redshift distribution is observed. The mean of redshift has drifted from about 2.8 in 2005 to about 2.1 by mid-2008 (see figure 1).
\newline 2. Over the same observation period we show that the mean time taken to acquire spectroscopic redshifts from the  GRB OA has become shorter. In the period 2005 to 2006 Nov we find that the average time to localize and acquire a spectroscopic redshift was about 800 min. In contrast, the response time in 2007 July to 2008 July has reduced to about 200 min. In 2008 large telescopes such as the VLT have managed to acquire OA localization and spectroscopic redshifts in under 100 min for half of the bursts observed.  
 \newline 3. There is a clear correlation between the mean time taken to acquire a spectroscopic redshift and the measured redshift. Specifically, shorter response times correspond to, on average, smaller redshifts (see Fig. 3). This is compelling evidence for a selection effect that biases optically bright high redshift bursts with longer response times. Conversely, for shorter response times, optically fainter bursts that are relatively closer will be bright enough for spectroscopic redshifts to be acquired.
 \newline 4. The improved efficiency in the response times to acquire redshifts is evidence that the learning curve effect is having an impact on the redshift statistics. Interestingly, the effect is not manifesting as a significant growth in the total number of measured redshifts over time. It does appear to be introducing a time-evolving bias for selecting OA at smaller redshifts (see Fig. 4) compared to earlier {\it Swift} detections prior to 2006. 

 It is interesting that the apparent deficit of $z<1.5$ bursts, evident in the redshift distribution prior to 2007 e.g. Coward et al. (2008), is less pronounced in 2007--2008. This could be a direct result of bursts in $z<1.5$ having relatively fainter OAs compared to higher redshift bursts. As the telescope response times have reduced, there are increasingly more fainter OA that are bright enough for spectroscopic measurement. We are currently investigating if the drift in the redshift distribution statistics can be used to probe the underlying reason why $z<1.5$ bursts have been relatively more difficult to determine. In our analysis, we have not attempted to construct a model to account for the shifting of average redshift measured with response time. In a follow-on study, we plan to investigate the correlations found in this study via simulation and by fitting a model to the averaged redshift-response time data. This will provide a means to confirm the claim in this study that GRBs at high-$z$ are intrinsically optically brighter relative to those bursts at low-medium redshift.
 
Finally, we highlight the critical role that telescopes capable of rapid response spectroscopy have on probing the environments of GRBs over different cosmological epochs. This work shows that to obtain an unbiased sample of redshifts, will require understanding a new bias that has not been considered previously, namely the response time to measure the redshift.     
 
\begin{figure} 
\includegraphics[scale=0.6]{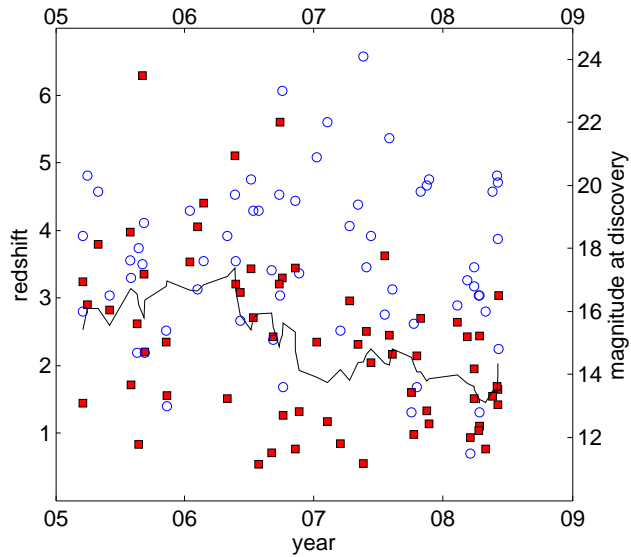}
\caption{The subset of 64 GRB redshifts with confirmed response times for spectroscopy (red squares) plotted as a time-series, starting in March 2005 and ending mid-2008. Output from a moving average filter for the same data is plotted (solid line), showing how the average redshift has a time-dependant trend towards decreasing values. The observed trend cannot be a result of evolution of the sources, but must be related to how the redshifts are determined. Optical magnitudes at discovery (blue circles) are also plotted, revealing a weak trend for observing increasingly brighter bursts at early times.  
 } \label{fig2} 
\end{figure}

\begin{figure} 
\includegraphics[scale=0.7]{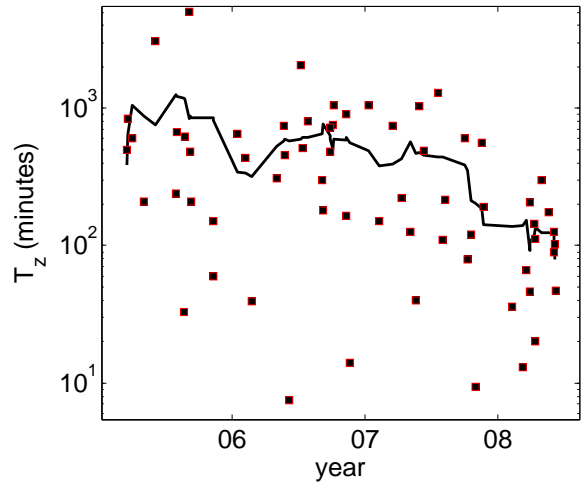}
\caption{ Plot showing the evolving reduction in average response times, $<T_z>$, to acquire a spectroscopic redshift from the GRB OA as a function of {\it Swift} mission time (solid line). The raw data (solid squares), taken from GCN circulars,  consists of 64 response times measured from the {\it Swift's} BAT trigger to the time of obtaining a high quality spectrum. The data span from 2005 March to mid 2008. Average response times have evolved from about 1000 min in 2005 down to several hundred minutes by mid-2008. This increase in efficiency is an indicator of the `learning curve' effect at work.  } \label{fig3} 
\end{figure}

\begin{figure} 
\includegraphics[scale=0.7]{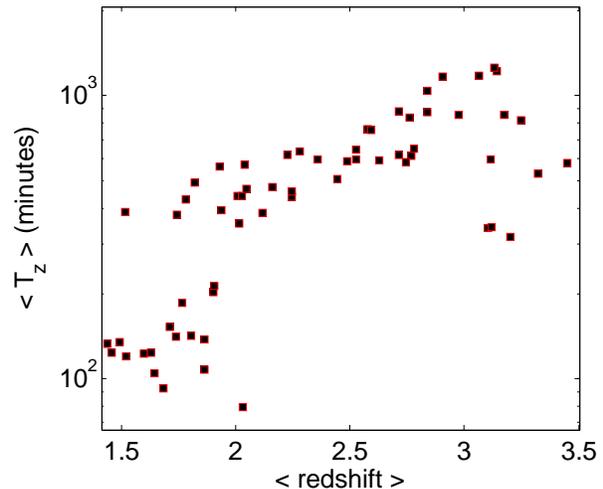}
\caption{ Plot showing the correlation of average response times, $<T_z>$, to acquire a spectroscopic redshift with the mean of the redshift distribution. The plot shows that the shortest response times on average correspond to smaller redshift bursts. This implies that these bursts have rapidly fading OAs that would have been too faint to acquire high quality spectroscopy from if observed at later times. Conversely, the longer duration response times may cause a preferential selection of  high-$z$ bursts that have very bright OAs and miss smaller-$z$ bursts if they have relatively fainter OAs. For $<T_z> \approx 1000$ min, the average redshift is about 3, and for $<T_z> \approx 200$ min, corresponding to observations in 2008 (see Fig. 2), the average redshift has reduced to about 2.} \label{fig3} 
\end{figure}

\begin{figure} 
\includegraphics[scale=0.7]{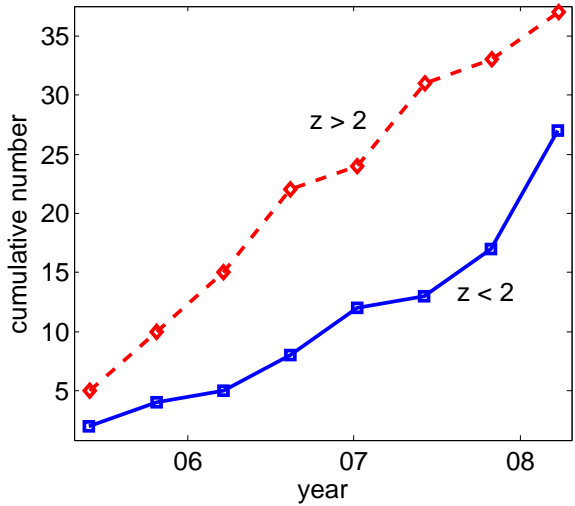}
\caption{ Plot showing the cumulative growth in measured redshifts over the time of the {\it Swift} mission for the two redshift regimes: $z<2$ (solid line) and $z>2$ (dashed line). The growth in $z>2$ redshifts is approximately linear while the numbers of $z<2$ redshifts is increasing at a marginally non-linear rate, mainly because of an enhancement in 2008.} \label{fig4} 
\end{figure}

\section*{Acknowledgments}
The author thanks Prof. D.G Blair and Dr R. Burman for useful comments and suggestions. D.M. Coward is supported by ARC grants DP0877550, LP0667494 and the University of Western Australia.
The author also thanks Dr M. Boer and Dr A. Klotz from the TAROT collaboration for providing positive feedback. Finally, the author thanks the referee for providing detailed comments that have significantly improved the clarity of the key findings.

\label{lastpage}

\end{document}